\documentclass[aps, prb, twocolumn, superscriptaddress, amsmath,  tightenlines, longbibliography]{revtex4-1}

\usepackage{dcolumn}
\usepackage{graphicx}
\usepackage{mathrsfs}
\usepackage{subfigure}
\usepackage{booktabs}
\usepackage{amsmath}
\usepackage{physics}
\usepackage{dsfont}
\usepackage{amstext}
\usepackage{amssymb}
\usepackage{amsbsy}
\usepackage{bbm}
\usepackage{amsthm}
\usepackage{graphicx}
\usepackage{color}
\usepackage[colorlinks,citecolor=blue]{hyperref}
\usepackage[left=1.3cm,   right=1.3cm,
top=1.20cm,    bottom=1.20cm]{geometry}

\usepackage{url}
\usepackage[colorlinks]{hyperref}
\hypersetup{%
	plainpages=true,
	breaklinks=true,       %not default in dvips mode, so we must specify
	hypertexnames=false,  %not ideal, but needed when pagenums duplicate (`i' vs. `1')
	pageanchor=true,
	colorlinks=true,
	linkcolor={blue},
	citecolor={red},
	urlcolor={blue},
	anchorcolor={black}
}

 \makeatletter

\newcommand{\Rmnum}[1]{\expandafter\@slowromancap\romannumeral #1@}
\makeatother

\begin{document}

\title{Localization-Delocalization Transitions  in Non-Hermitian Aharonov-Bohm Cages }
\author{Xiang Li}
\thanks{These authors contributed equally}
\affiliation{School of Physics and Optoelectronics, South China University of Technology,  Guangzhou 510640, China}
\author{Jin Liu}
\thanks{These authors contributed equally}
\affiliation{School of Physics and Optoelectronics, South China University of Technology,  Guangzhou 510640, China}
\author{Tao Liu}
\email[E-mail: ]{liutao0716@scut.edu.cn}
\affiliation{School of Physics and Optoelectronics, South China University of Technology,  Guangzhou 510640, China}

\date{{\small \today}}

%---------------------------------------------------------------------------

\begin{abstract}
  	A unique feature of non-Hermitian systems is the extreme sensitivity of the eigenspectrum to boundary conditions with the emergence of the non-Hermitian skin effect (NHSE). A NHSE originates from the point-gap topology of complex eigenspectrum, where an extensive number of eigenstates are anomalously localized at the boundary driven by nonreciprocal dissipation.  Two different approaches to create localization are disorder and flat-band spectrum, and their interplay can lead to the anomalous inverse Anderson localization, where the Bernoulli anti-symmetric  disorder induce mobility in a full-flat band system in the presence of  Aharonov-Bohm (AB) Cage.  In this work, we study the localization-delocalization transitions due to the interplay of the point-gap topology, flat band and correlated disorder in the one-dimensional rhombic lattice, where both its Hermitian and non-Hermitian structures show AB cage in the presence of magnetic flux. Although it remains the coexistence of localization and delocalization for the Hermitian rhombic lattice in the presence of the random anti-symmetric disorder, it surprisingly becomes complete delocalization, accompanied by the emergence of NHSE. To further study the effects from the Bernoulli anti-symmetric  disorder, we found the similar NHSE due to the interplay of the point-gap topology, correlated disorder and flat bands. Our anomalous localization-delocalization property can be experimentally tested in the classical physical platform, such as electrical circuit.
\end{abstract}

\maketitle

\section{Introduction} 

Recent years have witnessed tremendous advancements in exploring the physics of non-Hermitian Hamiltonian in open systems due to their  unique physical features without Hermitian counterparts \cite{RevModPhys.88.035002,PhysRevLett.116.133903, PhysRevLett.118.040401, PhysRevLett.118.045701,arXiv:1802.07964,El-Ganainy2018,ShunyuYao2018,PhysRevLett.125.126402,PhysRevLett.123.066404,Gao2015,PhysRevA.100.062118,PhysRevA.100.062131,Peng2014a,Peng2014b,PhysRevLett.117.110802,zdemir2019, YaoarXiv:1804.04672,PhysRevLett.121.026808,PhysRevLett.122.076801,Bliokh2019,PhysRevLett.123.170401, PhysRevLett.123.206404,PhysRevLett.123.066405,PhysRevLett.123.206404,PhysRevB.99.201103,Fan2021,Sun2021,Wu2021,Zhang2021, PhysRevB.100.054105,PhysRevB.99.235112,Zhao2019,PhysRevX.9.041015,PhysRevLett.124.056802,PhysRevB.102.235151, Li2020,PhysRevB.104.165117, Ashida2020, PhysRevB.102.205118,   PhysRevLett.124.086801, PhysRevLett.125.186802,PhysRevLett.127.196801,Li2021,Chen2022, RevModPhys.93.015005,PhysRevLett.128.223903,  Zhang2022,PhysRevLett.129.093001,Wang2023,Li2023,Zou2023, Ren2022,PhysRevX.13.021007,Lin2023,PhysRevLett.131.036402,PhysRevLett.131.116601, Okuma2023,arXiv:2311.06550,Leefmans2022,Leefmans2024,Zhang2024,Xie2024}.   The non-Hermitian Hamiltonian can be experimentally realized in  classical systems, such as  phononic lattice \cite{Zhou2023}, optical structures \cite{Weidemann2020,Wang2021}  and electrical circuits \cite{Helbig2020,Zou2021}, and open quantum systems in ultracold atoms \cite{PhysRevLett.129.070401, Ren2022}. Recent advances in non-Hermitian studies have shown many intriguing potential applications of the non-Hermitian systems  in  photonics \cite{feng2017non,el2018non,zdemir2019} and electrical circuits \cite{PhysRevLett.129.200201,PhysRevLett.130.077202}. One of  striking physical features in non-Hermitian systems is the extreme sensitivity of the eigenspectrum to boundary conditions with the emergence of the non-Hermitian skin effect (NHSE) \cite{ ShunyuYao2018, YaoarXiv:1804.04672,PhysRevLett.122.076801,PhysRevLett.123.066404,PhysRevLett.125.126402,  PhysRevLett.121.026808,Li2020,PhysRevB.104.165117, Ashida2020, PhysRevB.102.205118,   PhysRevLett.124.086801, PhysRevLett.125.186802}, where a large number of bulk modes  collapse into localized boundary modes in the open boundaries. The NHSE originates from the point-gap topology of the complex eigenspectrum, and has been shown to be responsible for many intriguing physical phenomena in non-Hermitian systems, such as the breakdown of conventional Bloch band theory \cite{ShunyuYao2018,PhysRevLett.123.066404,PhysRevLett.125.126402} and nonunitary scaling  of non-Hermitian localization \cite{PhysRevLett.126.166801}.

In contrast to the boundary localization of bulk modes caused by NHSE,  the disorder can induce Anderson localization of bulk modes along the lattice \cite{PhysRev.109.1492,RevModPhys.57.287}. Recent studies have shown that the introduction of  disorder into  non-Hermitian lattices with NHSE can lead to many unconventional phenomena \cite{PhysRevLett.77.570,PhysRevB.58.8384,PhysRevE.59.6433,   PhysRevX.8.031079, PhysRevB.100.054301, PhysRevB.101.165114, PhysRevLett.122.237601, PhysRevB.101.014202,PhysRevB.101.014204,Zhang2020,PhysRevB.103.L140201, PhysRevLett.126.090402, PhysRevB.104.104203,PhysRevLett.126.166801, PhysRevB.104.L121101, PhysRevLett.127.213601,Weidemann2021,Lin2022, PhysRevB.107.144204,arXiv:2311.03777, PhysRevB.109.L020202}, such as Anderson delocalization \cite{PhysRevLett.77.570,PhysRevX.8.031079}, nonunitary scaling rule  of non-Hermitian localization \cite{PhysRevLett.77.570} and reentrant NHSE \cite{arXiv:2311.03777}. In addition to the disorder,  flat bands can also lead to the localization of bulk modes along the lattice  due to the destructive interference among different propagation paths \cite{Leykam2018,Rhim2021}, providing another mechanism to confine waves.  A paradigmatic example of the flat-band localization is the Aharonov-Bohm (AB) caging, which provides the perfectly localized compact modes with   all the   bands being flat in, e.g., the one-dimensional (1D) rhombic lattice subjected to  an artificial
gauge field \cite{PhysRevLett.85.3906,PhysRevLett.88.227005, longhi2014aharonov,PhysRevLett.121.075502,Kremer2020,PhysRevLett.129.220403,Martinez2023}. Note that the interference among different propagation paths and AB caging  have been studied in higher dimension \cite{PhysRevLett.81.5888, PhysRevB.64.155306, PhysRevB.50.15953, PhysRevB.65.214504,PhysRevB.86.195403}. However, owing to the diverging effective mass in a flat band, the system becomes very sensitive to the disorder \cite{PhysRevLett.96.126401,PhysRevLett.116.066402}.  Remarkably, it has been shown that the  Bernoulli anti-symmetric  disorder  induces a localization–delocalization transition in the 1D rhombic lattice \cite{longhi2014aharonov,PhysRevLett.129.220403,Martinez2023}. This striking effect is dubbed the inverse Anderson transition from an insulating to a metallic phase, where the disorder removes geometric localization and restores transport in a lattice with all bands flat.

The interplay of non-Hermiticity and flat band  has recently received the extensive attentions \cite{PhysRevA.92.052103,PhysRevA.96.011802,PhysRevB.96.064305,PhysRevA.100.043808,PhysRevA.103.043329,PhysRevLett.123.183601,PhysRevLett.120.093901}. Its effects on the flat-band localization  in the 1D rhombic lattice have been also reported \cite{PhysRevResearch.2.033127,PhysRevA.108.023518,RamyaParkavi2022,PhysRevA.107.053508,	arXiv:2308.08418,MartnezStrasser2023}.  A natural question to ask is how the interplay of flat band, disorder and point-gap topology determines the localization-delocalization properties in the 1D rhombic lattice. In this work, we consider a  1D rhombic lattice subjected to the nonreciprocal hopping and magnetic flux, where we introduce two kinds of the correlated disorders, i.e., the random anti-symmetric disorder and the Bernoulli anti-symmetric  disorder.  In the Hermitian case, the Bernoulli anti-symmetric  disorder induces the inverse Anderson localization, while it leads to the NHSE in the presence of nonreciprocal hopping. Mostly interesting, the random anti-symmetric disorder causes the coexistence of localization and delocalization in the  non-Hermitian  rhombic lattice subjected to the magnetic flux, however, it induces the delocalization and the emergence of NHSE in the presence of the nonreciprocal hopping. 

The article is organized as follows.  In Sec.~\Rmnum{2},  we build the non-Hermitian model in the rhombic lattice, and study the flat-localization for the clean system in the presence of magnetic flux. In Sec.~\Rmnum{3}, we study the effects of random anti-symmetric disorder on the localization and delocalization property in the nonreciprocal model. In Sec.~\Rmnum{4}, we discuss the effects of Bernoulli anti-symmetric  disorder. In Sec.~\Rmnum{5}, we describe the experimental proposal for testing our theoretical results using electrical circuits.  In Sections \Rmnum{6}, we conclude the article.
	
\begin{figure}[!tb]
	\centering
	\includegraphics[width=9cm]{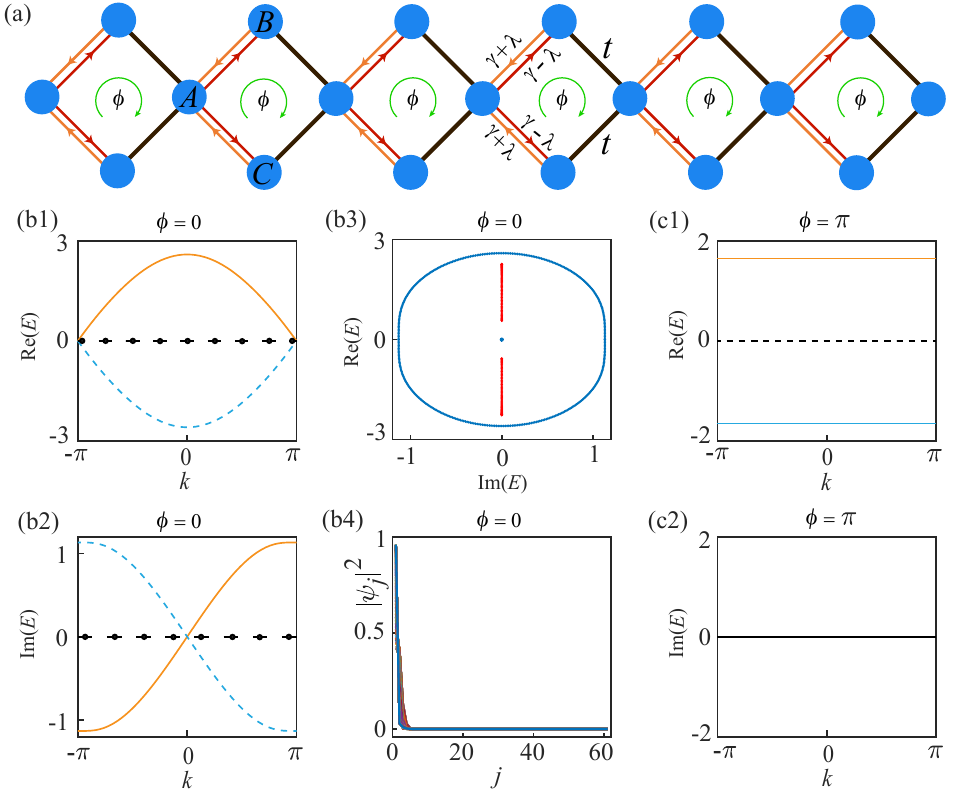}
	\caption{(a) Tight-binding representation of an asymmetric rhombic chain enclosed by a $\phi$ magnetic flux in each plaquette. Each unit cell contains three sublattices indicated by $A$, $B$ and $C$. $\gamma \pm \lambda$ denote the asymmetric hopping strengths (red and yellow lines with arrows), and $t$ is the symmetric hopping strength (black line). Real part (b1) and imaginary part (b2) of single-particle eigenspectrum for $\phi=0$. (b3) $\textrm{Re}(E)$ vs. $\textrm{Im}(E)$ of eigenenergies in complex plane with PBC (blue dots) and OBC (red dots) for $\phi=0$. (b4) Probability density distributions $\abs{\psi_j}^2$ (summed over each unit  cell) of eigenstates for their eigenenergies inside point gaps with $\abs{E} \neq 0 $ under OBC for $\phi=0$, where $\abs{\psi_j}^2 = \abs{\psi_{j,A}}^2 + \abs{\psi_{j,B}}^2 + \abs{\psi_{j,C}}^2$. Real part (c1) and imaginary part (c2) of single-particle eigenspectrum for $\phi=\pi$, and its bands are perfect flat, leading to mode localization. The other parameters are $\gamma/t = 1$ and $\lambda/t = 0.8$.}\label{lattice_model}
\end{figure}

\begin{figure}[!tb]
	\centering
	\includegraphics[width=9.0cm]{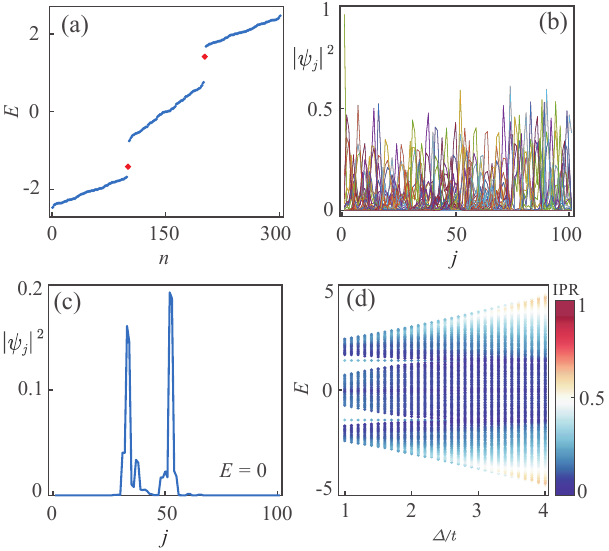}
	\caption{(a) Eigenenergies and (b) probability density distributions $\abs{\psi_j}^2$ (summed over each unit  cell) of eigenstates of the Hermitian rhombic lattice, subjected to the random anti-symmetric  disorder,  under OBCs for $\Delta / t=1$. The red dots indicate the topological boundary states. (c) $\abs{\psi_j}^2$ for $E=0$ with $\Delta / t=1$. (d) IPR vs. $\Delta$.  The other parameters for Hermitian conditions are $\phi=\pi$, $\gamma/t = 1$ and $\lambda/t = 0$.}\label{Fig21}
\end{figure}

\section{Model of Non-Hermitian rhombic lattice}

We consider a 1D rhombic lattice consisting of three coupled sublattices, indicated by $A$, $B$ and $C$  in Fig.~\ref{lattice_model}(a). In such a lattice, the asymmetric fermionic hopping within each unit cell is introduced, and a  magnetic flux with $U(1)$ Abelian gauge
fields  is applied to each plaquette. In the presence of  disordered onsite potential, the system's Hamiltonian is written as
\begin{align}\label{hamil1_0}
	\mathcal{H}_0   = & -t \sum_{j} \left(a_{j+1}^\dagger  b_{j} e^{i \phi} + a_{j+1}^\dagger  c_{j} + \textrm{H.c.}\right) \nonumber \\
	&  -(\gamma +\lambda )\sum_{j}     (a_{j}^\dagger  b_{j} + a_{j}^\dagger  c_{j})  \nonumber \\
	&  - (\gamma -\lambda ) \sum_{j} (b_{j}^\dagger  a_{j} + c_{j}^\dagger  a_{j}) \nonumber \\
	&  + \sum_{j,\alpha} \Delta^{(\alpha)}_{j} n_{\alpha, j},
\end{align}
where $a_{j}$, $b_{j}$ and $c_{j}$ is the annihilation operator at sublattices  $A$, $B$ and $C$ at  $j$th  unit cell, $\Delta^{(\alpha)}_{j}$ ($\alpha = A,B,C$) is the on-site disorder in   sublattice $\alpha$ at the $j$th unit cell, $n_{\alpha, j} = \alpha_{j}^\dagger \alpha_{j}$ ($\alpha = a,b,c$) denotes a density operator, $\gamma \pm\lambda$ represents the intracell asymmetric hopping strengths, $t$ is the intercell symmetric hopping strength. In the rhombic lattice, a single Peierls phase factor $\phi$ is used to represent the  magnetic flux  in each plaquette.

\begin{figure*}[!tb]
	\centering
	\includegraphics[width=18.6cm]{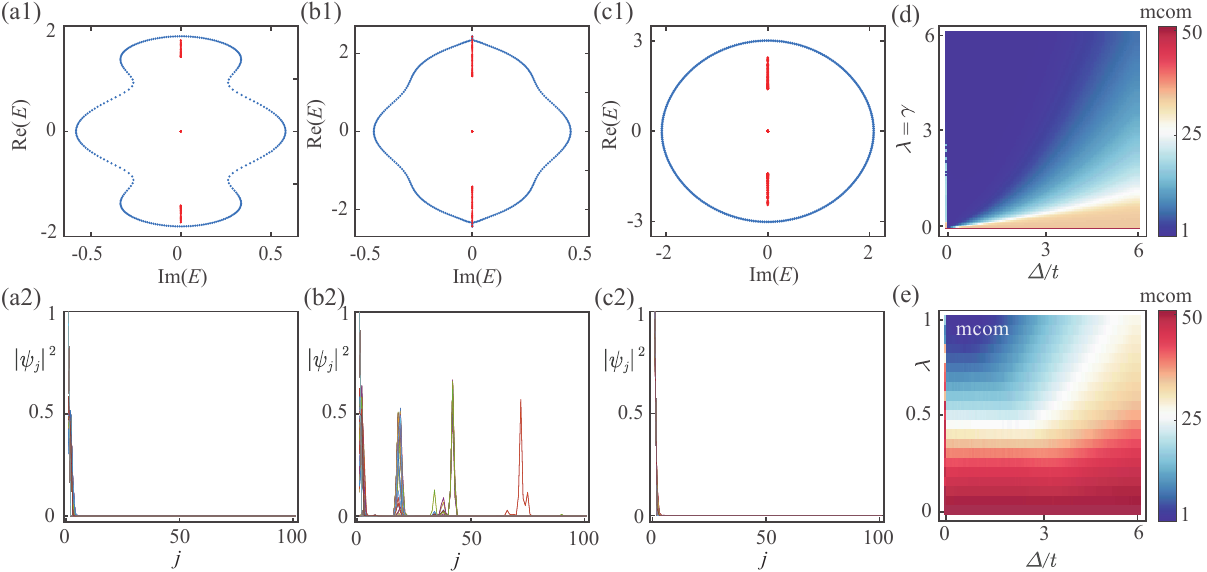}
	\caption{The localization and delocalization of  the non-Hermitian rhombic lattice subjected to random anti-symmetric  disorder $\Delta^{(B)}_{j}=-\Delta^{(C)}_{j}=\Delta_j$ ($\Delta_j \in [-\Delta/2,\Delta/2]$) for $\phi=\pi$. Complex eigenenergies under both OBCs (blue dots) and PBCs (red dots) (a1) for $\Delta / t=1$ and $\lambda/t=\gamma/t = 1$, (b1) for $\Delta / t = 2$ and $\lambda/t=\gamma/t = 1$, and (c1) for $\Delta / t=2$ and $\lambda/t=\gamma/t = 5$.  The corresponding probability density distributions $\abs{\psi_j}^2$ (summed over each unit  cell) of eigenstates are shown in (a2,b2,c2). (d) mcom as functions of $\lambda$ and $\Delta$ with $\lambda=\gamma$. (e) mcom as functions of $\lambda$ and $\Delta$ with  $\gamma/t = 1$. The mcom is averaged over 2000 disorder realization with $N=100$. }\label{Fig2}
\end{figure*}

\begin{figure}[!tb]
	\centering
	\includegraphics[width=8.9cm]{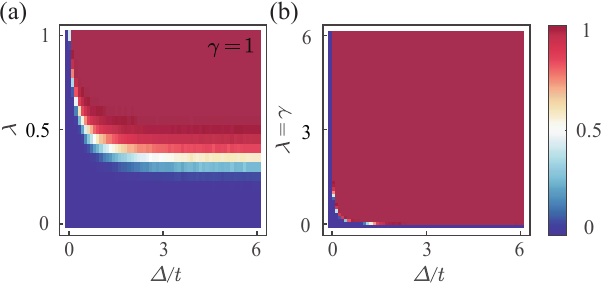}
	\caption{ (a) Winding number $w$ as functions of $\lambda$ and $\Delta$ with $\gamma/t = 1$ for the non-Hermitian rhombic lattice subjected to random anti-symmetric  disorder. (b) Winding number $w$ as functions of $\lambda$ and $\Delta$ with $\lambda = \gamma = 1$ for the non-Hermitian rhombic lattice subjected to random anti-symmetric  disorder. The results are averaged over $2000$ disorder realizations with $N=200$. }\label{Figadd}
\end{figure}

\begin{figure*}[!tb]
	\centering
	\includegraphics[width=18.0cm]{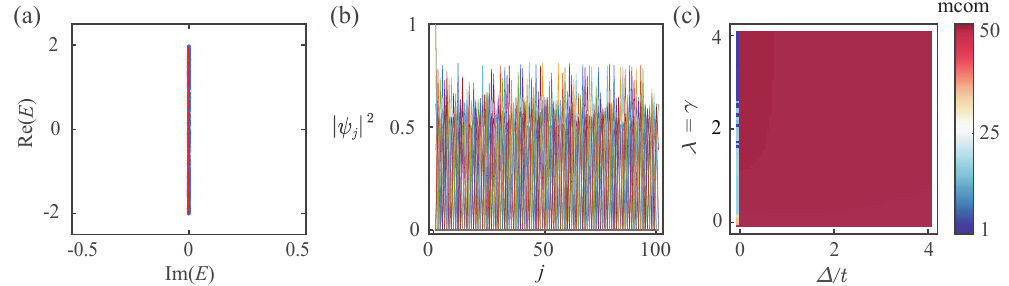}
	\caption{The localization  of  the non-Hermitian rhombic lattice subjected to random symmetric  disorder $\Delta^{(B)}_{j}=\Delta^{(C)}_{j}=\Delta_j$ ($\Delta_j \in [-\Delta/2,\Delta/2]$) for $\phi=\pi$. (a) Complex eigenenergies under both OBCs (blue dots) and PBCs (red dots) for $\Delta / t=1$, and $\lambda/t= \gamma/t = 1$. (b) The corresponding probability density distributions $\abs{\psi_j}^2$ (summed over each unit  cell) of eigenstates under OBCs.   (c) mcom as functions of $\lambda$ and $\Delta$ with $\lambda=\gamma$.    The mcom is averaged over 2000 disorder realization with $N=100$.}\label{Fig28}
\end{figure*}

In the absence of disorder in the system, i.e., $\Delta^{(\alpha)}_{j}=0$, we  plot the complex eigenenergies for $\phi=0$, as shown in Fig.~\ref{lattice_model}(b1-b3). There exist one flat band and two dispersive bands [see Fig.~\ref{lattice_model}(b1,b2)]. The eigenenergies with periodic boundary conditions (PBCs) form a point gap in the complex plane, and the eigenenergies with open boundary conditions (OBCs) lies inside the loop [see Fig.~\ref{lattice_model}(b3)]. A point gap usually indicates a non-Hermitian skin effect  for eigenstates \cite{PhysRevLett.124.086801}. Figure  \ref{lattice_model}(b4) show the probability density distribution $\abs{\psi_j}^2$, with $\abs{\psi_j}^2 = \abs{\psi_{j,A}}^2 + \abs{\psi_{j,B}}^2 + \abs{\psi_{j,C}}^2$, for $\abs{E} \neq 0 $, under OBC, where all modes with non-zero eigenenergies are localized at left boundaries. While zero-energy eigenstates are localized along the lattice due to the flat-band effect [see Fig.~\ref{lattice_model}(b1, b2).  The non-Hermitian skin modes can be characterized by the non-zero spectral winding number, defined as \cite{PhysRevLett.124.086801,PhysRevLett.125.126402}
\begin{align}\label{spectrum}
	\mathcal{W}(E_r)    = \frac{1}{2\pi i } \int_{0}^{2\pi} d k \partial_k \textrm{log}\, \textrm{det} [\mathcal{H}_0(k) - E_r] ,
\end{align}
where  $E_r$ is a chosen complex value as a reference energy, and $\mathcal{H}_0(k)$ is momentum-space Hamiltonian with
\begin{align}
	\mathcal{H}_0(k) &=\sum_{k} t \left[ e^{-i (k - \phi)}  a_{k}^\dagger  b_{k} + e^{-i k }  a_{k}^\dagger  c_{k} + \textrm{H.c.} \right] \notag \\
	&-(\gamma +\lambda ) \sum_{k} \left( a_{k}^\dagger  b_{k} +  a_{k}^\dagger  c_{k} \right) \notag \\
	&-(\gamma -\lambda ) \sum_{k} \left(   b_{k}^\dagger  a_{k} + c_{k}^\dagger a_{k}  \right).
\end{align}

For $\phi=\pi$, figure \ref{lattice_model}(c1,c2) shows three flat bands with perfectly compact localized states, the so-called Aharonov-Bohm cage \cite{PhysRevLett.81.5888,PhysRevLett.121.075502,Kremer2020,PhysRevLett.129.220403} in the non-Hermitian system.  In spite of nonreciprocal hopping, the destructive interference effect among different propagation paths greatly suppresses the non-Hermitian skin effect for $\phi = \pi$.

\section{Effects of random anti-symmetric  disorder}

We now consider effects of the interplay of  the nonreciprocal hopping and disorder on the mode localization and skin effects. We study two types of disorder realizations: (1)  random anti-symmetric  disorder, i.e,. $\Delta^{(B)}_{j}=-\Delta^{(C)}_{j}=\Delta_j$ with $\Delta_j$ sampled uniformly in the range $[-\Delta/2,~\Delta/2]$, and (2)   correlated binary (Bernoulli) anti-symmetric  disorder $\Delta^{(B)}_{j}=-\Delta^{(C)}_{j}=\Delta_j$, where $\Delta_j$ takes  two values $\pm \Delta$ with the same probability \cite{longhi2021inverse}. Unless specified otherwise, we assume $\phi=\pi$, and $\Delta^{(A)}_{j}= 0 $.

The strong random anti-symmetric disorder breaks the flatness of bulk bands, and  leads to a part of Anderson localization of eigenstates in the Hermitian rhombic lattice, as shown in Fig.~\ref{Fig21}(a,b). While, for the eigenenergies around zero, the eigenstates are extended [see Fig.~\ref{Fig21}(c)]. This indicates that disorder-induced
transport in the photonic cage system is possible. The 1D rhombic lattice supports topologically-protected in-gap states [see red dots in Fig.~\ref{Fig21}(a)], which has been experimentally observed \cite{Kremer2020}. 

To characterize the delocalization and localization induced by the random anti-symmetric disorder in a wide range of disorder strength $\Delta$, we calculate the inverse participation ratio (IPR) of each
normalized  right eigenstate $\psi_n = (\psi^{(a)}_n,\psi^{(b)}_n,\psi^{(c)}_n)^{T}$. The IPR is defined as
\begin{align}
	\textrm{IPR}_{n} =  \sum_{j} \left(| \psi_{n}^{(a)}(j)|^4  + | \psi_{n}^{(b)}(j)|^4   + | \psi_{n}^{(c)}(j)|^4  \right),
\end{align}
where the sums run over length $L$ of the rhombic chain, and $\sum_{j} \left(| \psi_{n}^{(A)}(j)|^2  + | \psi_{n}^{(B)}(j)|^2 +| \psi_{n}^{(C)}(j)|^2  \right) = 1$. If the $n$th eigenstate $\psi_{n}$ is extended, $\textrm{IPR}_{n} \simeq 1/(3L)$ and drops to zero for an infinite system. On the contrary, for the localized eigenstate $\psi_{n}$, $\textrm{IPR}_{n}$ keeps finite, and $\textrm{IPR}_{n} \simeq 1$ for the perfect localization.  Figure \ref{Fig21}(d) plots the eigenenergy-resolved IPR as a function of disorder strength $\Delta$. The random anti-symmetric disorder leads to the coexistence of localized and delocalized eigenstates in the Hermitian rhombic lattice.

We now proceed to study the effects of the random anti-symmetric disorder on the localization-delocalization properties of the  rhombic lattice in the presence of nonreciprocal hopping. Figures \ref{Fig2}(a1,a2), \ref{Fig2}(b1,b2) and \ref{Fig2}(c1,c2) plot the complex eigenenergies and the corresponding probability density distributions $\abs{\psi_j}^2$ (summed over each unit  cell) of eigenstates for different disorder strength $\Delta$ and unidirectional hopping strength $\lambda=\gamma$, respectively. In Fig.~\ref{Fig2}(a1,a2), using the same disorder strength as the one in  Fig.~\ref{Fig21}(a,b), the nonreciprocal hopping leads to the formation of the point gap under PBCs [see blue dots in Fig.~\ref{Fig2}(a1)], enclosing the eigenenergies under OBCs [see red dots in Fig.~\ref{Fig2}(a1)]. The point gap usually indicates the emergence of NHSE, where all the bulk modes are localized at the boundaries under OBCs [see Fig.~\ref{Fig2}(a2)]. The result shows that, although the  random anti-symmetric disorder causes the coexistence of localization and delocalization in the Hermitian lattice, the nonreciprocal hopping leads to the complete delocalization, accompanied by the NHSE. Therefore, the interplay of the flat band, disorder and point gap causes an unconventional localization-delocalization property in the  rhombic lattice.  Further increase  of the disorder strength $\Delta$ again leads to a part of localization and delocalization, where the skin modes and localized bulk states coexist [see \ref{Fig2}(b1,b2)]. While, the non-Hermitian skin effect reappears for the larger unidirectional hopping strength $\lambda=\gamma$, as shown in Fig.~\ref{Fig2}(c1,c2). These indicate that the interplay of random anti-symmetric disorder and nonreciprocal hopping can not only breaks the flatness of bulk bands, but also leads to complete delocalization, accompanied by the reentrant NHSE. 

To further explore the effects of asymmetrical hopping and disorder strength on the NHSE, we  calculate the average eigenstate localization in the form of the mean center of mass (mcom) of the amplitude squared of all eigenvectors $\psi_{n}$, averaged over the disorder realization \cite{PhysRevResearch.5.033058}, i.e.,
\begin{align}
	\mathrm{mcom}  = \frac{\sum_{j=1}^N j \left< \mathcal{A}(j) \right>_V}{\sum_{j=1}^N \left< \mathcal{A}(j) \right>_V} , 
\end{align}\label{eq:mcom}
with
\begin{align}
	\left< \mathcal{A}(j) \right>_V &= \left< \frac{1}{6N} \sum_{n=1}^N \left(| \psi_{n}^{(a)}(j)|^2  + | \psi_{n}^{(b)}(j)|^2 +| \psi_{n}^{(c)}(j)|^2  \right) \right>_V.
\end{align}
where, $\left< \cdot \right>_V$ indicates disorder averaging, and $N$ is the number of unit cells. The mcom determines the the degree of the mode localization. When mcom is close to be one or $N$, it indicates the most of eigenstates are localized at the boundaries with the emergence of non-Hermitian skin effect. Otherwise, it indicates the appearance of the delocalized eigenstates.

\begin{figure*}[!tb]
	\centering
	\includegraphics[width=18.8cm]{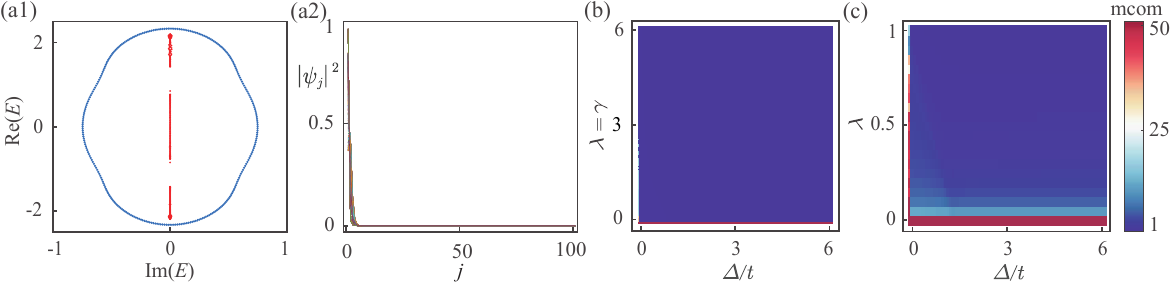}
	\caption{Localization and delocalization of  the non-Hermitian rhombic lattice subjected to Bernoulli anti-symmetric  disorder with $\Delta^{(B)}_{j}=-\Delta^{(C)}_{j}=\Delta_j$ ($\Delta_j$ randomly takes two values of $\pm\Delta$) for $\phi=\pi$. Complex eigenenergies under both OBCs (blue dots) and PBCs (red dots) (a1) for $\Delta / t=1$ and $\lambda/t=\gamma/t = 0.8$. The corresponding probability density distributions $\abs{\psi_j}^2$ (summed over each unit  cell) of eigenstates are shown in (a2). (b) mcom as functions of $\lambda$ and $\Delta$ with $\lambda=\gamma$. (c) mcom as functions of $\lambda$ and $\Delta$ with  $\gamma/t = 1$. The mcom is averaged over 2000 disorder realization with $N=100$. }\label{Fig3}
\end{figure*}

 \begin{figure*}[!tb]
 	\centering
 	\includegraphics[width=18cm]{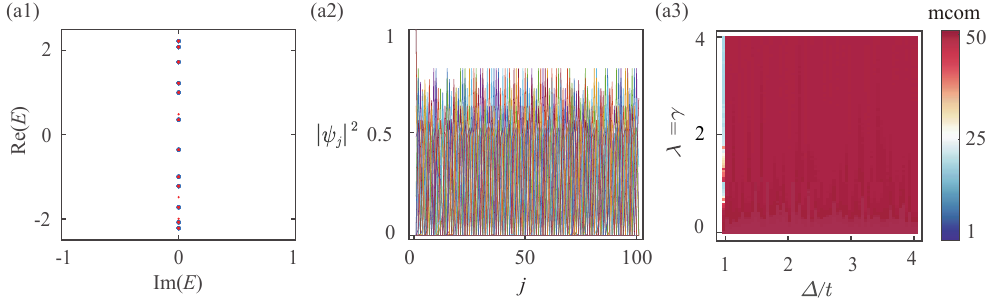}
 	\caption{Localization  of  the non-Hermitian rhombic lattice subjected to Bernoulli symmetric  disorder with $\Delta^{(B)}_{j}=-\Delta^{(C)}_{j}=\Delta_j$ ($\Delta_j$ randomly takes two values of $\pm\Delta$) for $\phi=\pi$ (a) Complex eigenenergies under both OBCs (blue dots) and PBCs (red dots) for $\Delta / t=1$, and $\lambda/t= \gamma/t = 1$. (b) The corresponding probability density distributions $\abs{\psi_j}^2$ (summed over each unit  cell) of eigenstates under OBCs f.   (c) mcom as functions of $\lambda$ and $\Delta$ with $\lambda=\gamma$.     The mcom is averaged over 2000 disorder realization with $N=100$. 
 	}\label{Fig5add}
 \end{figure*}

Figure \ref{Fig2}(d) plots the mcom as functions of $\lambda$ and $\Delta$ with $\lambda=\gamma$. When the asymmetrical hopping strength $\lambda=\gamma$ is fixed, the increasing disorder strength $\Delta$ leads to the state localization. While, as the $\lambda=\gamma$ rises, the localized states become skin modes, i.e., the emergence of NHSE induced by random anti-symmetric  disorder. Moreover, the appearance of NHSE requires a strong asymmetric hopping for the fixed values of $\gamma$ and $\Delta$, as shown in  Fig.~\ref{Fig2}(e), where we plot the mcom as functions of $\lambda$ and $\Delta$ with  $\gamma/t = 1$.

To characterize the  topology of non-Hermitian skin effects, we calculate the real-space winding number, which is defined as \cite{PhysRevB.103.L140201}
 \begin{align}
 	w(E_b) = \frac{1}{L'} \text{Tr}' \left( \hat{Q}^{\dagger} [\hat{Q},\hat{X}] \right),
 \end{align}
where $\hat{Q}$ is positive-definite Hermitian matrix, which is obtained by the polar decomposition $(\mathcal{H}-E_b) = \hat{Q}\hat{P}$, with unitary matrix $\hat{P}$. $\hat{Q}$ and $\hat{P}$ are related to singular value decomposition $(\mathcal{H}_0-E_b) = \hat{M}\hat{S}\hat{N}^\dagger$, with $\hat{Q} = \hat{M}\hat{N}^\dagger$ and $\hat{P} = \hat{N}\hat{S}\hat{N}^\dagger$. $\hat{X}$ is the coordinate operator, with $X_{jj',ss'}= j\delta_{j,j'}\delta_{s,s'} (s=a,b)$, and $\text{Tr}'$ denotes the trace over the middle interval with length $L'$, where the whole chain is cut off from both ends. This definition of winding number avoids the effects from the system's boundary.  
 
Figure \ref{Figadd}(a) plots the winding number $w$ as functions of $\lambda$ and $\Delta$ with $\gamma/t = 1$ for the non-Hermitian rhombic lattice subjected to the random anti-symmetric  disorder. The result shows that the emergence of non-Hermitian effects is strongly determined by the asymmetric strength $\gamma$. While, for the perfect unidirectional hopping with $\lambda=\gamma$, even though the disorder is very strong, there exists skin modes [see Fig.~\ref{Figadd}(b)]. These skin modes are coexistence with the localized modes for the regime with the large   mcom [see Fig.~3(b1,b2,d)].

Although the  random anti-symmetric disorder leads to the delocalization in the nonreciprocal rhombic lattice subjected to the $\pi$ gauge field, the random symmetric disorder with $\Delta^{(B)}_{j}=\Delta^{(C)}_{j}=\Delta_j$ ($\Delta_j \in [-\Delta/2,\Delta/2]$) leads to the Anderson localization, as shown in Fig.~\ref{Fig28}(a,b), where there doesn't exist a point gap with the absence of NHSE for $\Delta / t=1$, and  $\lambda/t= \gamma/t = 1$. We calculate the mcom as functions of $\lambda$ and $\Delta$ with $\lambda=\gamma$ in Fig.~\ref{Fig28}(c). Indeed, when the nonreciprocal rhombic lattice is subjected to the random symmetric disorder,   all the states remain localized, and there is no NHSE.

\section{Effects of Bernoulli anti-symmetric  disorder }

The random anti-symmetric disorder cause the  coexistence of localized and delocalized states in the Hermitian rhombic lattice subjected to the $\pi$ flux. while, the interplay of the  random anti-symmetric disorder, flat band and point gap leads to the delocalization, accompanied by the reentrant NHSE.  In the Hermitian rhombic lattice, it has been shown that the Bernoulli anti-symmetric  disorder leads to the inverse Anderson localization due to the interplay of  geometric frustration and disorder \cite{longhi2021inverse}.

We now consider the non-Hermitian rhombic lattice subjected to the Bernoulli anti-symmetric  disorder with $\Delta^{(B)}_{j}=-\Delta^{(C)}_{j}=\Delta_j$ ($\Delta_j$ randomly takes two values of $\pm\Delta$) for $\phi=\pi$. Figure \ref{Fig3}(a1) shows the complex eigenenergies of the lattice  under both OBCs and PBCs. The point gap (blue dots) enclosing the real eigenvalues (red dots) under OBC indicate the NHSE, as shown in Fig.~\ref{Fig3}(a2), where all the bulk modes are localized at the left boundary. To explore how the asymmetrical hopping and disorder strength influence the NHSE, we calculate the mcom as functions of $\delta=\gamma$ and $\Delta$, as shown in Fig.~\ref{Fig3}(b). In contrast to the case of random anti-symmetric disorder, a small value of the unidirectional hopping strength $\delta=\gamma$ can induced the NHSE in spite of the disorder strength. Moreover, a small degree of asymmetric hopping can cause the skin modes in spite of the disorder strength [see Fig.~\ref{Fig3}(c)].

 Although the  Bernoulli anti-symmetric disorder leads to the delocalization in the nonreciprocal rhombic lattice subjected to the $\pi$ gauge field, the Bernoulli symmetric disorder with $\Delta^{(B)}_{j}=\Delta^{(C)}_{j}=\Delta_j$ ($\Delta_j$ randomly takes two values of $\pm\Delta$) leads to the Anderson localization, as shown in Fig.~\ref{Fig5add}(a,b), where there doesn't exist a point gap with the absence of NHSE for $\Delta / t=1$, and  $\lambda/t= \gamma/t = 1$. We calculate the mcom as functions of $\lambda$ and $\Delta$ with $\lambda=\gamma$ in Fig.~\ref{Fig5add}(c). Indeed, when the nonreciprocal rhombic lattice is subjected to the random symmetric disorder,   all the states remain localized, and there is no NHSE.

\section{Experimental proposal}

\begin{figure*}
	\centering
	\includegraphics[width=18cm]{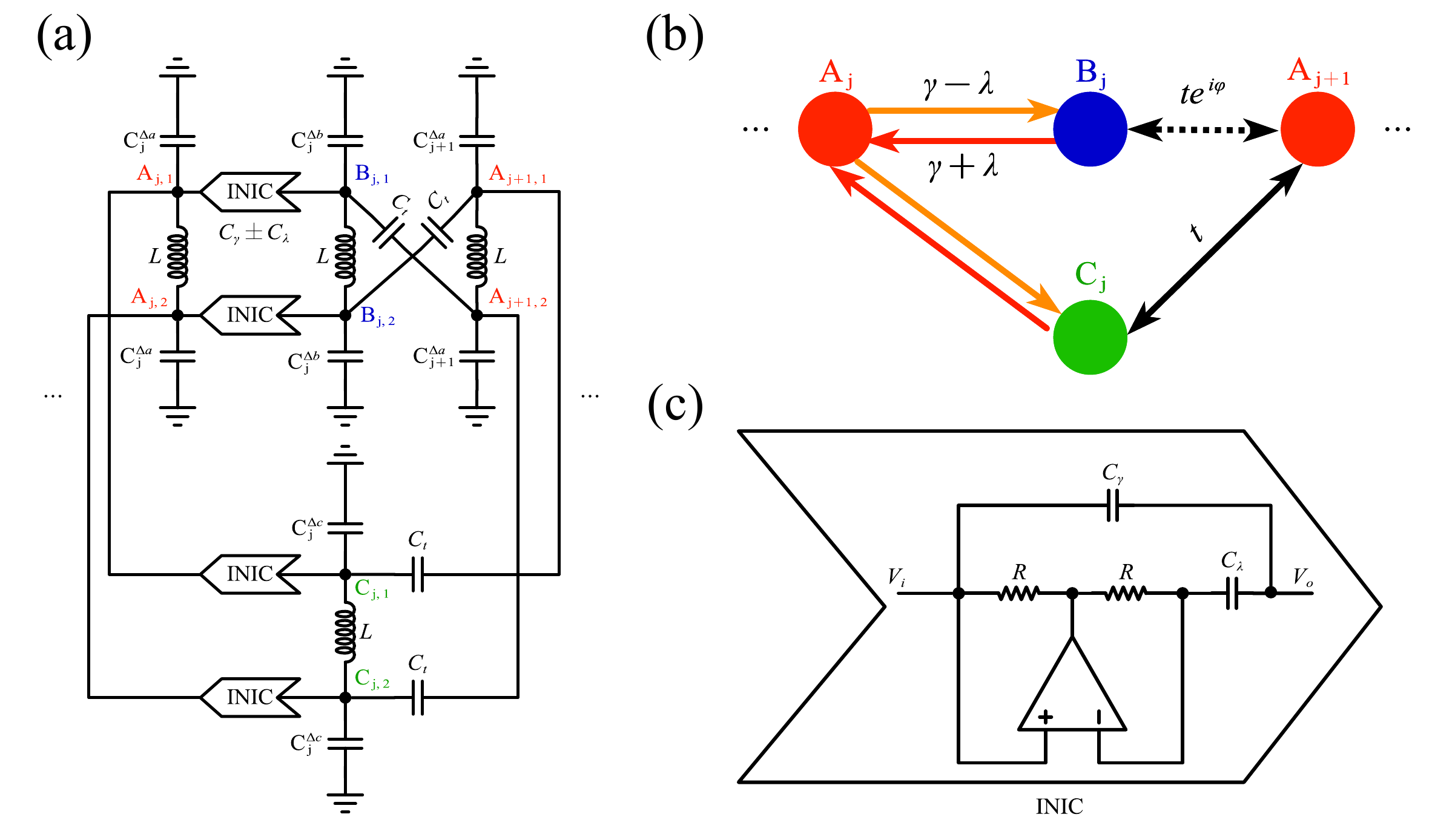}
	\caption{(a) Electrical circuit implementation of the model in Eq.~(\ref{hamil1_0}),  corresponding to the lattice structure in (b). The nonreciprocal hopping between nodes $j$ and $j+1$ is realized by the negative impedance converters through current inversions (INICs).  (c) Details of INIC. }
	\label{fig1}
\end{figure*}

The  localization-delocalization transition induced by the anti-symmetric disorder with the emergence of NHSE can be experimentally observed in the electrical circuits \cite{Helbig2020,Zou2021}. we design non-Hermitian electrical circuits, corresponding to the model in Eq.~(\ref{hamil1_0}), as shown Fig.~\ref{fig1}. The nonreciprocal hopping between nodes $j$ and $j+1$ is realized by the negative impedance converters through current inversions (INICs) \cite{PhysRevLett.122.247702}.  The model in Eq.~(\ref{hamil1_0}) is represented by the circuit Laplacian $J(\omega)$ of the circuit. The Laplacian is defined as the response of the grounded-voltage vector  $\mathbf{V}$ to the vector $\mathbf{I}$ of input current   by\cite{dong2021topolectric,lee2018topolectrical}   
\begin{align}\label{eq2}
	\mathbf{I}(\omega) = J(\omega)\mathbf{V}(\omega).
\end{align}

In Fig.~\ref{fig1}, the negative impedance converter through circuit reads $C_\gamma \pm C_\lambda$, introducing the nonreciprocal intracell hopping in Eq.~(\ref{hamil1_0}). The grounded capacitors, $C_j^{\Delta a}$, $C_j^{\Delta b}$ and $C_j^{\Delta c}$, represents the on-site disorder of $A$, $B$ and $C$ sublattices in the $j$th unit cell. Capacitor $C_t$ is used to represent the symmetrical intercell hopping of the model in Eq.~(\ref{hamil1_0}). The phase $\phi=\pi$ can be achieved by crossing the adjacent nodes with wires  in Fig. \ref{fig1}(a). The inductor $L$ is used to tune the resonant frequency of the circuit. Using Eq.~(\ref{eq2}), the current of each node within the unit cell can be expressed as
\begin{widetext}
\begin{align}\label{eq31}
		I_{a, j}=~&i\omega \left( C_{\gamma}+C_{\lambda} \right) \mathcal{I}_2V_{b, j}+i\omega C_t\mathcal{I}_2V_{c, j-1}+i\omega C_t\left( \begin{matrix}
			0&		1\\
			1&		0\\
		\end{matrix} \right) V_{b, j-1}+i\omega \left( C_{\gamma}+C_{\lambda} \right) \mathcal{I}_2V_{c, j}+\frac{1}{i\omega L}\left( \begin{matrix}
			-1&		1\\
			1&		-1\\
		\end{matrix} \right) V_{a, j} \nonumber \\
		&-i\omega \left( 2C_{\gamma}+2C_{\lambda}+2C_t+C_{j}^{\Delta a} \right) \mathcal{I}_2V_{a, j},
\end{align}
	\begin{align}\label{eq32}
					I_{b, j}=i\omega \left( C_{\gamma}-C_{\lambda} \right) \mathcal{I}_2 V_{a, j}+i\omega C_t\left( \begin{matrix}
				0&		1\\
				1&		0\\
			\end{matrix} \right) V_{a, j+1}+\frac{1}{i\omega L}\left( \begin{matrix}
				-1&		1\\
				1&		-1\\
			\end{matrix} \right) V_{b, j}-i\omega \left( C_{\gamma}-C_{\lambda}+C_t+C_{j}^{\Delta b} \right) \mathcal{I}_2 V_{b, j},
	\end{align}
\begin{align}\label{eq33}
			I_{c, j}=i\omega \left( C_{\gamma}-C_{\lambda} \right) \mathcal{I}_2 V_{a, j}+i\omega C_t\mathcal{I}_2 V_{a, j+1}+\frac{1}{i\omega L}\left( \begin{matrix}
			-1&		1\\
			1&		-1\\
		\end{matrix} \right) V_{c, j}-i\omega \left( C_{\gamma}-C_{\lambda}+C_t+C_{j}^{\Delta c} \right) \mathcal{I}_2 V_{c, j},
\end{align}
\end{widetext}
where the vectors $I_{\alpha, j} = (I_{\alpha, 1,j},~I_{\alpha, 2,j})^T ~(\alpha=a, b, c)$ and $V_{\alpha, j} = (V_{\alpha, 1,j},~V_{\alpha, 2,j})^T~(\alpha=a, b, c)$ denote  the node currents  and voltages   of $A$, $B$ and $C$ sublattices within the $j$th unit cell, respectively, and $\mathcal{I}_2$ represents the $2 \times 2$ identity matrix.

We transform Eqs.~(\ref{eq31})-(\ref{eq33})  using a unitary matrix  
\begin{align}\label{eq4}
	U=\frac{1}{\sqrt{2}}\left( \begin{matrix}
		1&		1\\
		1&		-1\\
	\end{matrix} \right),
\end{align}
, and achieve the transformed current-voltage relationship  
\begin{widetext}
\begin{align}\label{eq51}	
		\bar{I}_{a, j}=~&i\omega \left( C_{\gamma}+C_{\lambda} \right) \mathcal{I}_2\bar{V}_{b, j}+i\omega C_t\mathcal{I}_2\bar{V}_{c, j-1} +i\omega C_t\left( \begin{matrix}
			1&		0\\
			0&		e^{i\pi}\\
		\end{matrix} \right) \bar{V}_{b, j-1} +i\omega \left( C_{\gamma}+C_{\lambda} \right) \mathcal{I}_2\bar{V}_{c, j}+\frac{1}{i\omega L}\left( \begin{matrix}
			0&		0\\
			0&		-2\\
		\end{matrix} \right) \bar{V}_{a, j} \nonumber \\
		&-i\omega \left( 2C_{\gamma}+2C_{\lambda}+2C_t+C_{j}^{\Delta a} \right) \mathcal{I}_2\bar{V}_{a, j}, 
\end{align}
\begin{align}\label{eq52}
		\bar{I}_{b, j}&=i\omega \left( C_{\gamma}-C_{\lambda} \right) \mathcal{I}_2\bar{V}_{a, j}+i\omega C_t\left( \begin{matrix}
			1&		0\\
			0&		e^{i\pi}\\
		\end{matrix} \right) \bar{V}_{a, j+1} +\frac{1}{i\omega L}\left( \begin{matrix}
			0&		0\\
			0&		-2\\
		\end{matrix} \right) \bar{V}_{b, j}-i\omega \left( C_{\gamma}-C_{\lambda}+C_t+C_{j}^{\Delta b} \right) \mathcal{I}_2\bar{V}_{b, j},
	\end{align}
\begin{align}\label{eq53}
			\bar{I}_{c, j}&=i\omega \left( C_{\gamma}-C_{\lambda} \right) \mathcal{I}_2\bar{V}_{a, j}+i\omega C_t\mathcal{I}_2\bar{V}_{a, j+1} +\frac{1}{i\omega L}\left( \begin{matrix}
			0&		0\\
			0&		-2\\
		\end{matrix} \right) \bar{V}_{c, j}-i\omega \left( C_{\gamma}-C_{\lambda}+C_t+C_{j}^{\Delta c} \right) \mathcal{I}_2\bar{V}_{c, j}.
\end{align}
\end{widetext}
As shown in Eqs.~(\ref{eq51})-(\ref{eq53}),   after the unitary transformation, the current-voltage equations at second node is decoupled with ones at first node, and the phase $e^{i\pi}$ of the intercell hopping  appears in the second node of the circuit. Therefore, we only need to consider the circuit Laplacian at second node.   We rewrite the current-voltage equations for the second nodes as
\begin{widetext}
\begin{align}\label{eq61}
		\bar{I}_{a, 2, j}=~&i\omega \left( C_{\gamma}+C_{\lambda} \right) \bar{V}_{b, 2, j} +i\omega C_t \bar{V}_{c, 2, j-1} +i\omega C_te^{i\pi}\bar{V}_{b, 2, j-1} +i\omega \left( C_{\gamma}+C_{\lambda} \right) \bar{V}_{c, 2, j} -\frac{2}{i\omega L}\bar{V}_{a, 2, j}  \nonumber \\
		&-i\omega \left( 2C_{\gamma}+2C_{\lambda}+2C_t+C_{j}^{\Delta a} \right) \bar{V}_{a, 2, j}, 
\end{align}
\begin{align}\label{eq62}
		\bar{I}_{b, 2, j}&=i\omega \left( C_{\gamma}-C_{\lambda} \right) \bar{V}_{a, 2, j} +i\omega C_te^{i\pi}\bar{V}_{a, 2, j+1} -\frac{2}{i\omega L}\bar{V}_{b, 2, j} -i\omega \left( C_{\gamma}-C_{\lambda}+C_t+C_{j}^{\Delta b} \right) \bar{V}_{b, 2, j},
\end{align}
\begin{align}\label{eq63}
		\bar{I}_{c, 2, j}&=i\omega \left( C_{\gamma}-C_{\lambda} \right) V_{a, 2, j}^{*}+i\omega C_t\bar{V}_{a, 2, j+1} -\frac{2}{i\omega L}\bar{V}_{c, 2, j} -i\omega \left( C_{\gamma}-C_{\lambda}+C_t+C_{j}^{\Delta c} \right) \bar{V}_{c, 2, j},
\end{align}
\end{widetext}
where $\bar{I}_{\alpha, 2, j} ~(\alpha=a, b, c)$ and $\bar{V}_{\alpha, 2, j} ~(\alpha=a, b, c)$ denote  the current and voltage of the second node within each sublattice in circuit, respectively. Therefore, we achieve the targeted circuit Laplacian $J(\omega)$ to simulate the   model in Eq.~(\ref{hamil1_0}) as
\begin{widetext}
\begin{equation}
	\mathcal{J}=i\omega 
		\begingroup
		\setlength{\tabcolsep}{4pt}             % Default value: 6pt
		\renewcommand{\arraystretch}{2}        % Default value: 1	
		\begin{pmatrix}
			\eta_{\Delta a}  &		C_{\gamma}+C_{\lambda}&		C_{\gamma}+C_{\lambda}&		\cdots&		0&		0 &		0 \\			C_{\gamma}-C_{\lambda}&	\eta_{\Delta b}&		0&		\cdots&		0&		0&		0\\
			C_{\gamma}-C_{\lambda}&		0&		\eta_{\Delta c}&		\cdots&		0&		0&		0\\
			\vdots&		\vdots&		\vdots&		\ddots&		\vdots&		\vdots&		\vdots\\
			0&		0&		0&		\cdots&		\eta_{\Delta a} &		C_{\gamma}+C_{\lambda}&		C_{\gamma}+C_{\lambda}\\
			0&		0&		0&		\cdots&		C_{\gamma}-C_{\lambda}&		\eta_{\Delta b}&		0\\
			0&		0&		0&		\cdots&		C_{\gamma}-C_{\lambda}&		0&		\eta_{\Delta c}  \\
		\end{pmatrix}.
		\endgroup	
\end{equation}
\end{widetext}
where $\eta_\alpha = \frac{2}{\omega ^2L}-\left(M+C_{j}^{\alpha} \right)$ ($\alpha = \Delta a, \Delta b, \Delta c$) with $M=2C_{\gamma}+2C_{\lambda}+2C_t$, $N=C_{\gamma}-C_{\lambda}+C_t$.

\section{Conclusion}

In this work, we study how the nonreciprocal hopping determines the localization and delocalization  properties in the 1D rhombic lattice subjected to the  magnetic flux and correlated disorder. When the Bernoulli anti-symmetric  disorder is introduced into the non-Hermitian rhombic lattice, it leads to the boundary localization of the bulk modes in spite of the disorder strength. To be interesting, a small degree of asymmetric hopping can cause the NHSE.  When the random anti-symmetric  disorder is introduced into the non-Hermitian rhombic lattice, it leads to the anomalous delocalization, accompanied by the NHSE, while the random anti-symmetric disorder causes the coexistence of localization and delocalization in the Hermitian rhombic lattice. Moreover, the localization-delocalization transition strongly depends on the disorder strength and asymmetric hopping strength. The experimental setup for observing the effects of the point-gap, flat band correlated disorder is proposed in electrical circuits.

\begin{acknowledgments}
T.L. acknowledges the support from the Fundamental Research Funds for the Central Universities (Grant No.~2023ZYGXZR020), Introduced Innovative Team Project of Guangdong Pearl River Talents Program (Grant No.~2021ZT09Z109), and the Startup Grant of South China University of Technology (Grant No.~20210012).
\end{acknowledgments}

%\bibliography{Reference}


\end{document}